# OPERATIONAL ASPECTS OF THE HIGH CURRENT UPGRADE AT THE UNILAC

J. Glatz, J. Klabunde, U. Scheeler, D. Wilms, GSI, Darmstadt, Germany

*Abstract*

With the new GSI High Current Injector, the beam pulse intensity will be increased by more than two orders of magnitude. The high beam power and the short stopping range at particle energies below 12 MeV/u can destroy accelerator components even during a single beam pulse. Therefore, the operation of the whole accelerator facility has required major changes in hardware, software and operating strategy. A sophisticated beam diagnostic system is indispensable for a safe operation. Preferably non-destructive devices were installed. Destructive elements, e.g. beam stoppers, slits, apertures, were improved in order to withstand the high beam power. Automatic damage prevention was realised by beam loss monitors comparing and evaluating very fast beam current transformer signals. Additionally, the component status will be controlled permanently. For foil stripping at 11.4 MeV/u, a magnetic beam sweeping system was installed Thereby, the hit area will be increased during one 100 µs pulse. During operation, manual variation of parameters has to be reduced. Set-up and automatic beam adjustment procedures have to exclude uncontrolled beam loss.

The versatility of the UNILAC is enhanced by the possible three-beam operation on a pulse-to-pulse basis.

Since November 1999 the upgraded UNILAC is serving the experiments.

## 1 INTRODUCTION

With the installation of the High Current Injector HSI, the beam pulse power has been increased considerably. The schematic layout of the UNILAC is shown in Fig. 1. where key parameters for the high current acceleration are listed, uranium being the reference ion. Considering the high beam pulse power ( maximum of 1250 kW at the gas stripper section ) and especially the short stopping range at the UNILAC energies, accelerator components could be destroyed. Even one pulse with a length of 100 µs can melt metal surfaces. Necessary consequences for a safe operation will be discussed in the following sections, the topics are: control of the high intensity beam, safety of accelerator components, foil stripping at high beam power, operating strategy including the time sharing operation.

The stepwise installation and commissioning of the new linac sections were carried out from January to November 1999. There are separate reports concerning the commissioning of the high current ion sources, LEBT, HSI, and the new stripper section at 1.4 MeV/u. The summary report on the commissioning contains an extensive list of references. (Ref. [1]).

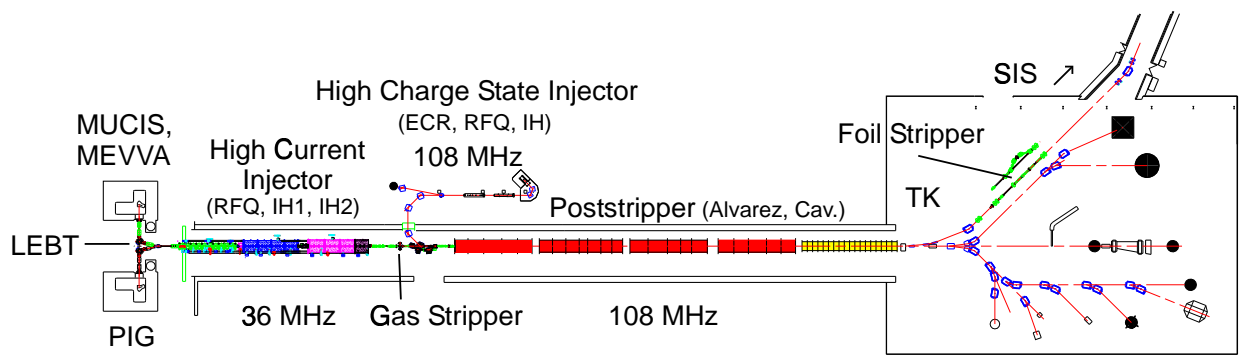

| | | | |
|---|---|---|---|
| Energy | 0.0022 MeV/u | 1.4 MeV/u | 11.4 MeV/u (for SIS Injection) |
| | | | 3.6 - 14 MeV (for UNILAC Exp.) |
| Charge State | 4+ | 4+ / 28+ | 28+ / 73+ |
| Intensity | ≥ 16 emA | 15 emA / 12.6 emA | 12.6 emA / 5.0 emA |
| Beam Pulse Power | 2.1kW | 1250 kW / 150 kW | 1220 kW / 183 kW |
| Stopping Range | ~ 0.03 µm | ~ 5 µm | ~ 50 µm |

Fig. 1: Schematic layout of the UNILAC with some key parameters, uranium as reference ion

## 2 CONTROL OF THE HIGH INTENSITY BEAM

### 2.1 Beam Diagnostics

Considering the beam power and the short stopping range, all diagnostic elements had to be replaced preferably by non-destructive devices. Beam current is measured by means of beam transformers instead of Faraday cups, beam position with segmented capacitive pick-ups and secondary ion beam monitors instead of profile harps. Descriptions and specifications of these elements are given in a separate contribution (Ref.[2]).

Along the UNILAC, 33 beam transformers are installed capable to measure pulse current in the range from 500 nA to 30 mA.

At 24 positions improved capacitive pick-ups are installed. They are used to measure intensity, width and phase of bunches. By evaluation of pick-up signals at different positions, energy and longitudinal emittance can be estimated. By subdivision of the ring electrode into 4 sectors the horizontal and vertical positions are determined.

For on-line beam profile measurements, residual gas ionisation monitors has been developed. Due to its size (length is 250 mm), a limited number (at present 8) of such elements are installed at important matching points.

### 2.1 Beam Stoppers, Slits, Apertures

For safe operation improved beam stoppers, slits, and apertures were installed at certain positions. In Fig. 2 the increase of the temperature is shown, if an intense beam hits a tungsten plate. In a very thin layer the temperature is high. Therefore, the exposed area of a beam stopping device had to be increased.

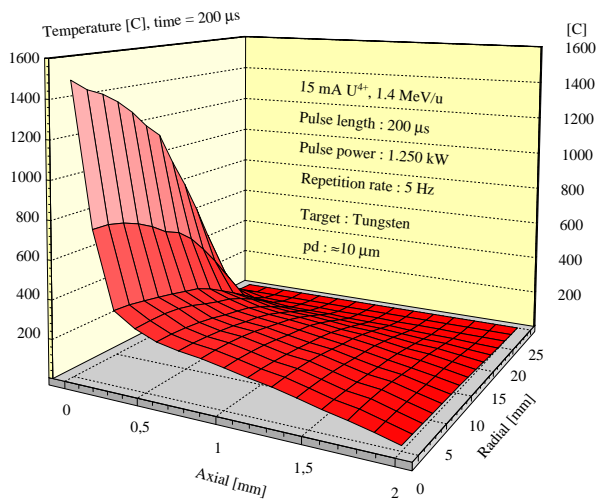

Figure 2: Uranium ions at 1.4 MeV/u hit a tungsten plate

### 2.3 Interlock System

An active protection of accelerator components is realised by fast beam transformer measurements. Along the linac, beam loss which can destroy hardware will be measured at different key positions The principle of beam loss control is shown in Fig. 3. The integral current of two beam transformers will be compared. If the difference is above a critical level of charge, the beam pulse length will be reduced with the chopper in the LEBT. The system consisting of 10 stations (later 20) was successfully tested. At some positions the current signals are distorted by fields of magnets, a better shielding is necessary.

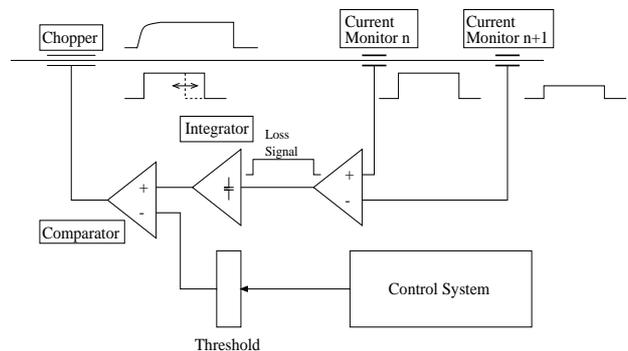

Fig. 3: Principle of beam loss control

In addition to the automatic survey, a component status control is permanently active. If failures occur, the beam will be switched off by the chopper or beam stoppers. The status will be controlled by software which delivers detailed information about the nature of component failure.

## 3 FOIL STRIPPING AT HIGH CURRENT

In order to achieve the highest ion energies in the synchrotron, the charge state must be as high as possible. Therefore, carbon foils are used, the thickness ranges from 100 to 600 $\mu g/cm^2$. The primary intensity from the UNILAC to fill the SIS up to the space charge limit is about $10^{11}$ ions per pulse. Due to the absorption of up to 3 % of beam energy, a fast heating of the foil arises. The deposited energy density is decreased by increasing the effective beam cross section by two steps ( Fig. 4 ):

- Enlarged vertical beam size of 20 mm diameter.
- Fast horizontal sweeping of the beam on the foil (about 50mm) during the beam pulse by two symmetrical sweeper magnets in combination with a quadrupole.

Additionally, pulsed magnets allow to deflect a beam of low intensity to a second foil. Without deflection, a beam without stripping could be also transported to the SIS. By that the flexibility for serving the physics experiments with the SIS is considerably increased, because time sharing of these three modes can be implemented.

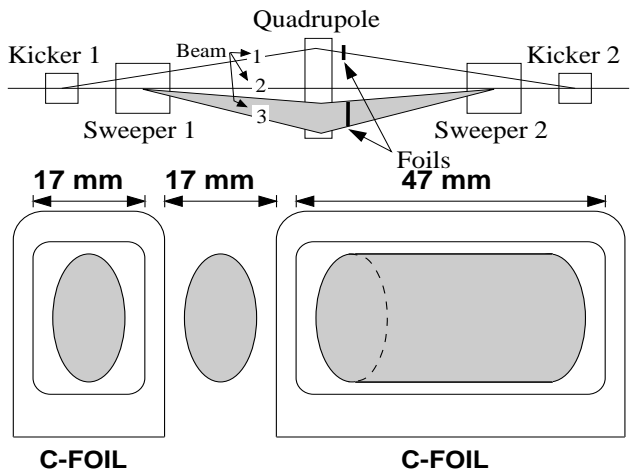

Fig. 4: Schematic drawing of the foil stripper section

## 4  OPERATING STRATEGY

For the accelerator operation at high current, the manual variation of operational parameters was reduced drastically to avoid uncontrolled beam handling and loss. Experimentally confirmed setting values for all components of the linac must be provided. Furthermore, for set-up and adjustment procedures the beam will be reduced by a suitable shortening of the pulse length. After the fine tuning of the linac, the access to the machine parameters are blocked. The development of a new operating software which includes particularities of high current acceleration is underway. The automatic calculation and setting of current dependent magnet set-values are already implemented. An on-line control of beam position using the signals of capacitive pick-up probes will be available soon.

## 5  TIME SHARING OPERATION

In the past fast switching of ions on a pulse-to-pulse basis was possible only in the poststripper linac by selecting the ions either from the HLI or the former Wideröe prestripper linac. All magnets of the injectors could be operated only in a dc mode. The new injector allows time sharing between ion beams from the PIG and the MUCIS/MEVVA terminal. All new magnets of the LEBT, rf linac, and the quadrupoles and dipoles of the stripper section can be pulsed. Now an arbitrary time sharing operation with up to three beams is possible. The allowed beam data of the three terminals are listed in Table 1. The enlarged versatility of the GSI facility is offered now routinely. An example of a three-beam operation for four different experiment stations is shown in Fig. 5.

Table 1: Limits of beam parameters of the three terminals

| Terminal | Ion Source | Mass/Charge | Frequency | Pulse Length |
|---|---|---|---|---|
| HLI | ECR | 8.5 | 50 Hz | 7 ms |
| RIGHT | PIG | 26 | 50 Hz | 7 ms |
| LEFT | MUCIS/MEVVA, | 65 | 5 Hz | 0.3 ms |
|  |  | 65 | 10 Hz | 0.2 ms |
| RIGHT | PIG | 65 | 16.7 Hz | 1.2 ms |

## 6  SUMMARY

Since November 1999 the new injector is used for serving physics experiments. With argon from the MUCIS, a new intensity maximum was attained. First successful operation at high beam power levels confirmed the applied strategy. In future we expect that the available tools allow a successful and safe operation of uranium at the highest design level of beam power.

## REFERENCES

[1] W. Barth, Commissioning of the 1.4 MeV/u High Current Heavy Ion Injector, these Proceedings.
[2] P. Forck, , Measurement of the 6-dimensional Phase Space at the New GSI High Current Linac, these Proceedings.

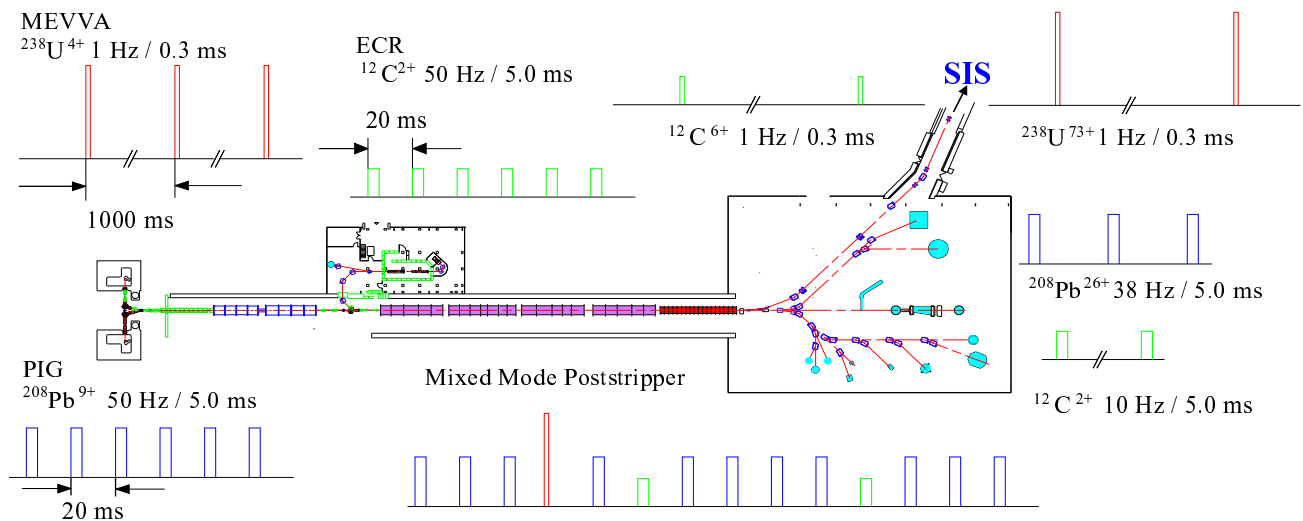

Fig. 5: Example of a three-beam operation